\documentclass[graybox]{svmult}

\usepackage{mathptmx}       
\usepackage{helvet}         
\usepackage{courier}        
\usepackage{type1cm}        
\usepackage{makeidx}         
\usepackage{graphicx}        
\usepackage{multicol}        
\usepackage[bottom]{footmisc}

\usepackage{natbib}

\def\spose#1{\hbox to 0pt{#1\hss}}
\def\lta{\mathrel{\spose{\lower 3pt\hbox{$\mathchar"218$}}
     \raise 2.0pt\hbox{$\mathchar"13C$}}}
\def\gta{\mathrel{\spose{\lower 3pt\hbox{$\mathchar"218$}}
     \raise 2.0pt\hbox{$\mathchar"13E$}}}

\makeindex             
\def\aj{AJ}

\def\apj{ApJ}
\def\apjl{ApJL}

\def\aap{A\&A}

\def\mnras{MNRAS}

\def\nat{Nature}

\begin{document}

\title*{The Galactic Bar}
\author{Ortwin Gerhard and Christopher Wegg}
\institute{Ortwin Gerhard, Christopher Wegg \at Max Planck Institute for extraterrestrial Physics, PO Box 1312, Giessenbachstr.,
   85741 Garching, Germany, \email{gerhard@mpe.mpg.de, wegg@mpe.mpg.de}}
%
%
\maketitle

\abstract{The Milky Way's bar dominates the orbits of stars and the
  flow of cold gas in the inner Galaxy, and is therefore of major
  importance for Milky Way dynamical studies in the Gaia era. Here we
  discuss the pronounced peanut shape of the Galactic bulge that has
  resulted from recent star count analysis, in particular from the VVV
  survey. We also discuss the question whether the Milky Way has an
  inner disky pseudo-bulge, and show preliminary evidence for a
  continuous transition in vertical scale-height from the peanut
  bulge-bar to the planar long bar.}

\section{Introduction}
\label{sec:intro}

The Gaia satellite will soon provide us with exquisite data for the
motions of stars in our Galaxy. The Milky Way's stellar bar has a strong
influence on the orbits of stars inside the solar circle, and
dominates the gas flow in the inner Galaxy. To understand the
structure and dynamics of the Galactic bar better is therefore of
prime importance in the Gaia era.

Renewed interest in the Galactic bar and bulge has also been triggered
recently by unprecedented new near-infrared (NIR) photometric and
spectroscopic surveys of the inner Galaxy.  In this contribution, we
summarize recent work on the structural properties of the Galactic bar
and bulge based on such survey data.

About two thirds of all large disk galaxies are barred
\citep{Eskridge+00}, including our Milky Way. Historically, several
perceptive papers already suggested in the 1970s that the non-circular
motions seen in HI observations towards the inner Galaxy could
naturally be interpreted in terms of elliptical gas streamlines
\citep{Shane72,Peters75}. But it was not until the 1990s that the
barred nature of the Milky Way bulge was considered established. This
revised view of our Galaxy was based on new data and ideas at that
time: the NIR photometry by the COBE satellite
\citep{Weiland+94,Binney+97}, enabling a global view of the bulge
through the dust layer; new analysis of the inner Milky Way gas flows
traced in longitude-velocity diagrams \citep{Englmaier+Gerhard99,
  Fux99}, in the context of triaxiality and bars in external galaxies;
asymmetries in bulge star counts
\citep{Stanek+94,Lopez-Corredoira+97}, and the newly measured optical
depth for microlensing towards bulge fields
\citep{Paczynski+94,Bissantz+97}. These data, as well as the
kinematics of bulge stars available at the time \citep{Zhao+94,Fux97},
could all be consistently explained in terms of a barred bulge model
with its major axis in the first quadrant. 2MASS NIR star counts
\citep{Lopez-Corredoira+05,Skrutskie+06}, the cylindrical rotation in
the bulge \citep{Kunder+12,Ness+13b,Shen+10}, and other data have
since confirmed and refined our understanding of the Galactic bar.

In external disk galaxies seen at low inclination, bars are easy to
see but their vertical structure is not well-constrained. However, we
now know from gas- and stellar-kinematic data that box or
peanut-shaped (b/p) bulges in edge-on systems are related to bars
\citep{Merrifield+Kuijken99,Chung+Bureau04}. Photometric data
\citep{Luetticke+00,Bureau+06} are consistent with predictions from
N-body models that bars in galaxies generally consist of an inner
three-dimensional (3D) part, the b/p bulge, and a more extended 2D
part of the bar \citep{Athanassoula05}.  In these simulations, a bar
first grows from an unstable disk and then becomes unstable to a
buckling instability which leads to an inner boxy bulge
\citep{Combes+90, Raha+91, Athanassoula+Misiriotis02}. The bar can
then grow further through angular momentum loss to the dark matter
halo, and eventually go through a second buckling instability which
leads to the formation of a strongly peanut-shaped bulge
\citep{Martinez-Valpuesta+06}. Similar secular evolution has been
observed also in cosmologically more realistic simulations of disk
galaxies, such as dissipative collapse models
\citep{Samland+Gerhard03,Ness+14}, or hierarchical disk galaxy models
in which strong feedback and the absence of significant mergers ensure
a relatively quiescent star formation history
\citep{Martig+12,Guedes+13}.

\section{Bulge distances from K-band magnitudes of red clump giants}
\label{sec:rcg}

Red clump giants are immensely useful for star count studies of the
Galaxy. They are He core burning stars with a narrow range of
luminosities and colours, hence can be employed to map out the
distant-dependent structure of the Galactic bulge.  Using the
$\alpha$-enhanced BASTI isochrones for various metallicities at age 10
Gyr from \citet{Pietrinferni+04}, a metallicity-averaged luminosity
function of bulge red clump (RC) stars has Gaussian $\sigma(K)=0.18$
mag, $\sigma(J-K)=0.05$, and $M_{K,RC}=-1.72$ in the $K_S$-band
\citep{Wegg+Gerhard13}.  The $K_S$-band magnitude is slightly brighter
than the value from solar-neighbourhood RC stars \citep{Alves00}. In
the colour-magnitude diagram, RC stars are spread because of
variations in distance, reddening, age, and metallicity.  The
variations in absolute $K_S$-magnitude with age and metallicity are
$\Delta_{\rm age} M_{K,RC}\sim 0.03$/Gyr at age 10 Gyr and $\Delta_Z
M_{K,RC}\sim-0.28$/dex \citep{Salaris+Girardi02}. From the same study,
for old stellar populations the number of RC stars per unit initial
mass varies by $\sim 10\%$ for metallicities $\gta 0.02$
solar. Therefore the distribution of RC stars is a good tracer of the
stellar mass distribution in the bulge. Distances to individual RC
giants in the upper bulge can be estimated with $\sigma_D\sim
10\%$. At low latitudes, residual dust extinction broadens the clump
magnitude distribution, leading to $\sigma_D\sim 15\%$ at $|b|=1\deg$
\citep{Gerhard+Martinez-Valpuesta12}.

\section{The Galactic box/peanut bulge}
\label{sec:bpbulge}
The COBE data firmly established the boxy shape and the presence of
longitudinal asymmetries in the Galactic bulge, which appears brighter
and vertically more extended on the $l>0$ side, but symmetric in
latitude $b$ \citep{Weiland+94}. These asymmetries are a signature of
a triaxial bulge whose major axis is tilted in the Galactic plane
relative to the line-of-sight (LOS) from the Sun to the Galactic
centre \citep{Blitz+Spergel91}.  From these asymmetries, parametric
and non-parametric bulge luminosity models were reconstructed
\citep{Dwek+95,Bissantz+Gerhard02} and then used to study the
influence of the Galactic bar on the gas dynamics and microlensing
observations.  However, uncertainties remained due to the lack of
distance information in the COBE data and to the uncertainties in the
extinction maps.

Red clump star counts in several fields across the bulge first
illustrated the different distances to the near and far sides of the
triaxial bulge \citep{Stanek+94,Stanek+97}. With the 2MASS all sky
data a first star count reconstruction of the bulge shape became
possible, even if somewhat affected by the limit in the survey depth
\citep{Lopez-Corredoira+05}. More recent analysis of OGLE-III and
2MASS RC counts in upper bulge fields near the minor axis showed the
presence of two density peaks along the LOS
\citep{Nataf+10,McWilliam+Zoccali10}. These results were interpreted
as an X-shape in the bulge density distribution such as seen in N-body
models of peanut bulges. Abundances and velocities derived from
spectra of large numbers of bulge stars showed that only the more
metal-rich stellar populations participate in this 'split red clump'
\citep{Ness+12}.

The recent VVV survey is $\sim3-4$ mag deeper than 2MASS and RC stars
are visible all through the bulge even quite close to the Galactic
plane.  These data have brought a quantum leap for constraining the
bulge density distribution. First results at $b=\pm1\deg$ showed
clearly the in-plane tilt of the barred bulge, and also a structural
change in the central $|l|<4\deg$ \citep{Gonzalez+11b}, indicating a
less elongated distribution of stars in this region
\citep{Gerhard+Martinez-Valpuesta12}.  In our recent work
\citep{Wegg+Gerhard13}, we used the VVV DR1 data \citep{Saito+12a} for
some $300$ sightlines across $|l|<10\deg$ and $-10\deg<b<5\deg$, to
measure the 3D density distribution of the Galactic bulge. For all
these LOS, we constructed extinction- and completeness-corrected
K$_{\rm S}$-band magnitude distributions, and deconvolved these with
the K-band luminosity function of RC giants and red giant branch bump
(RGBB) stars. The extinction model was determined from the RC stars
themselves, with a method very similar to
\citet{Gonzalez+11b}. Completeness effects were corrected based on
artificial stars inserted in the VVV images.  For the deconvolution we
used a Lucy-Richardson scheme including a 'background component' of
stars not in the RC+RGBB. This is a good approximation because the
magnitudes of these stars vary over a wide range.  The deconvolved LOS
density distributions clearly show the split red clump at latitudes
$|b|\ge 4\deg$.

\begin{figure}[htbp]
\centering
\includegraphics[scale=0.95]{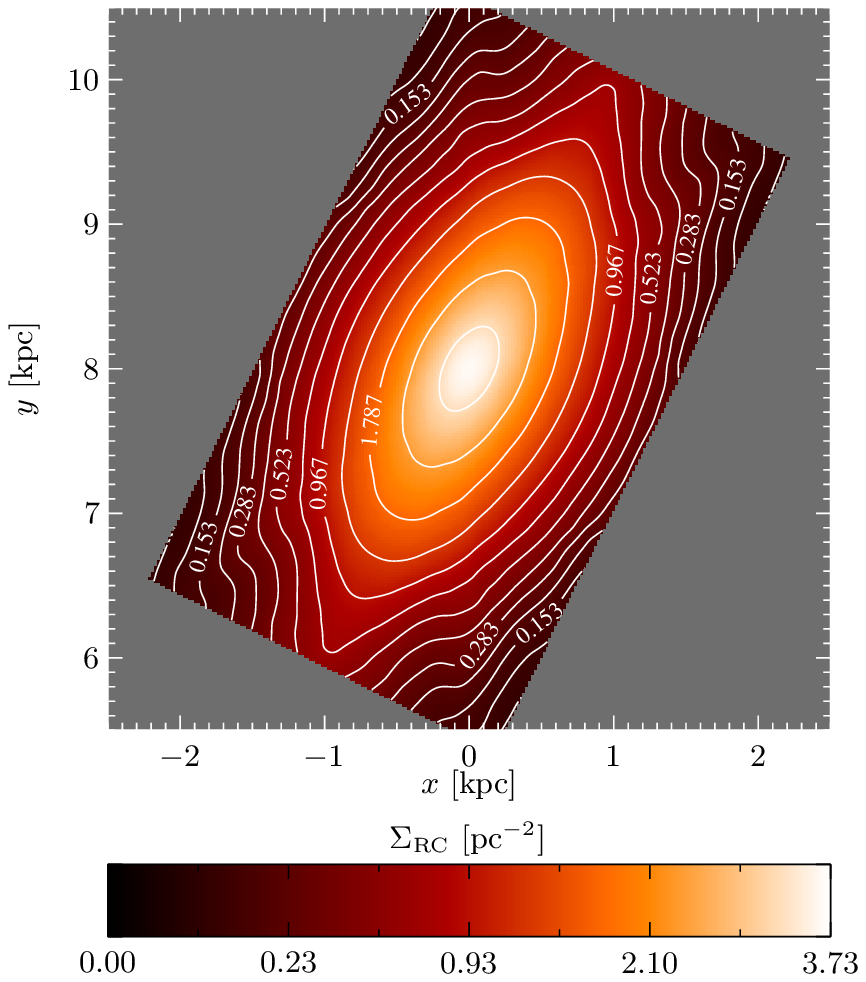}
\includegraphics[scale=0.76]{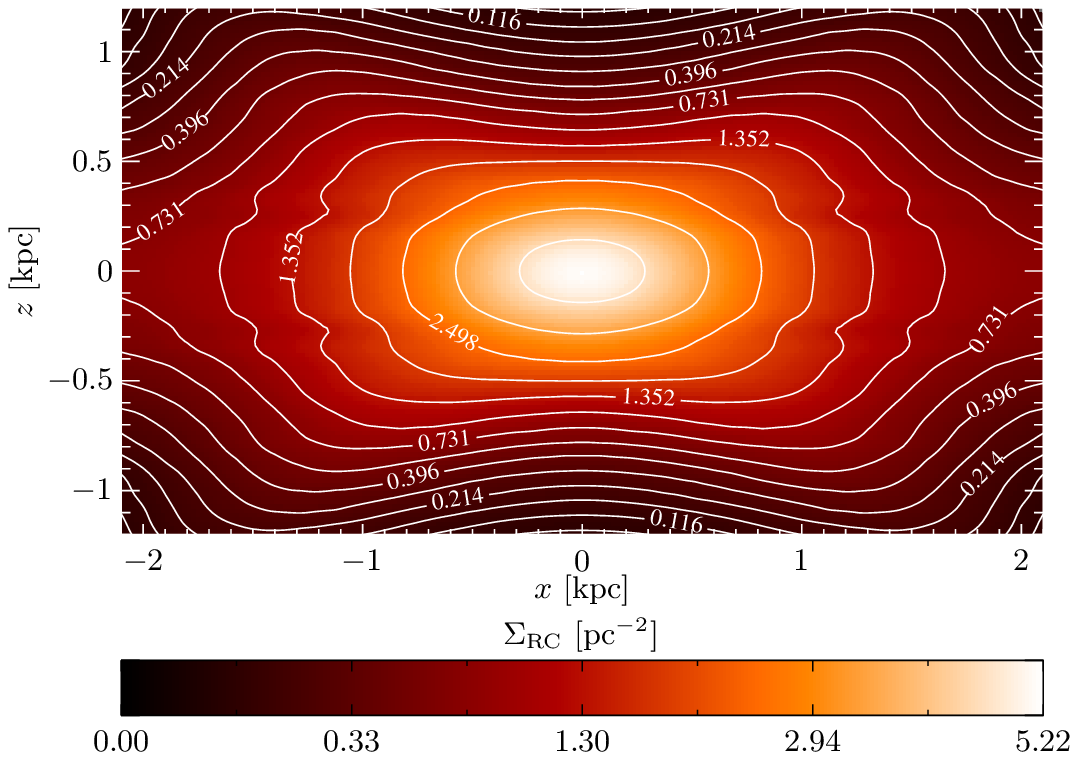}
\caption{{\bf (a), top:} The Galactic bulge as viewed from the
  north Galactic pole.  {\bf (b), bottom:} The bulge 3D density
  projected along the intermediate axis. Numbers give the surface
  density of RC stars in pc$^{-2}$, contours are spaced by 1/3
  mag. The 3D density map extends to 1.15 kpc along the minor axis and
  to 1.4 kpc along the intermediate axis, and the projections are
  within these limits. }
\label{fig:1}
\end{figure}

For constructing the entire 3D bulge density distribution, we then
assumed that the bulge is eightfold triaxially symmetric. This
assumption can be checked at the end: the rms deviations between the
mirror-symmetric points in the final model are indeed small. Finding
the best eightfold symmetric distribution determines the tilt angle of
the bar-bulge to the LOS to be $(27\pm2)\deg$, where the dominant
error is systematic, arising from the details of the deconvolution
process. Our final density map covers the inner
($2.2\times1.4\times1.1$) kpc of the bulge/bar.  It is illustrated in
Figures~\ref{fig:1} and \ref{fig:2}.  The density is typically
accurate to $\sim10\%$ except at $y>1$ kpc along the intermediate axis
of the bar. In the projected density maps shown in Figs.~\ref{fig:1},
the region $|b|<1\deg$ was extrapolated from $b\ge1\deg$ with a simple
sech$^2$-model (Portail et al., in preparation), because at these low
latitudes the extinction and completeness corrections are too large
for a reliable density deconvolution.

Fig.~\ref{fig:1} shows a highly elongated bar, with projected axis
ratios $\approx(1:2.1)$ for isophotes reaching major axis radii $\sim
2$ kpc. Above about 400 pc above the Galactic plane, a prominent
X-structure is visible which explains the earlier measurements of the
split red clump. This X-structure is characteristic for a class of b/p
bulges \citep{Bureau+06}. In fact, the b/p bulge of the Milky Way is
very strong, about as strong as in the proto-typical b/p bulge in NGC
128 \citep{Wegg+Gerhard13b}. Beyond about 1 kpc along the major axis,
the iso-surface density contours in the edge-on projection bend down
and outwards to the Galactic plane.

\section{Does the Milky Way have a disky pseudo-bulge?}

\begin{figure}[htbp]
\centering
\includegraphics[scale=.85]{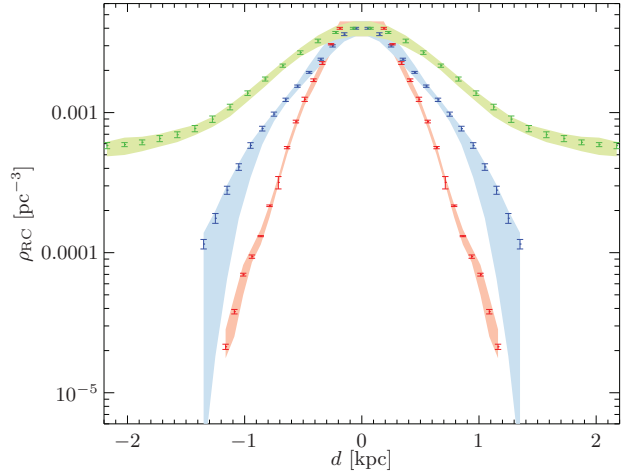}
\caption{Density of RC stars along the major axis (green),
  intermediate (blue), and minor axis (red) for the bulge. The major
  and intermediate axis profiles are offset from the Galactic plane by
  187.5 pc. The error bars are the rms deviations between the
  eightfold symmetric points of the final bulge density
  distribution. The shaded regions show estimated systematic errors
  based on varying the details of the deconvolution procedure.  From
  \citet{Wegg+Gerhard13}.}
\label{fig:2}
\end{figure}

Along the principal bar axes the density falls of approximately
exponentially, as shown in Fig.~\ref{fig:2}.  Exponential
scale-lengths are (0.70:0.44:0.18) kpc, corresponding to axis ratios
(10:6.3:2.6). The flattening of the major axis profile beyond
$\sim 1$ kpc signifies the transition to the longer, planar bar.

Particularly noteworthy is the very short vertical scale height in the
centre, perhaps indicative of a central disk-like, high-density
pseudo-bulge structure as is seen in many early and late type b/p
bulge galaxies \citep{Bureau+06}, including the edge-on Milky
Way analogue NGC 4565 \citep{Kormendy+Barentine10}.

Additional support for this interpretation comes from the
change-of-slope at $|l|\simeq 4\deg$ seen in the VVV RC star count
maxima longitude profiles for $b=\pm1\deg$ \citep{Gonzalez+11b,
  Wegg+Gerhard13}. The simplest interpretation of this result is the
presence of a rounder, more nearly axisymmetric central part of the
bar, as illustrated by \citet{Gerhard+Martinez-Valpuesta12} with an
N-body model with this property (see Figure 3). In fact, if the central parts of the
Milky Way bar were completely axisymmetric, the inner slope of the
longitude profile would be nearly zero. This model also predicted that
the transition to a rounder inner parts was confined to a few degrees
from the Galactic plane. This was later confirmed with VVV data by
\citet{Gonzalez+12}.  However, more work is required particularly at
low latitudes to confirm a disky pseudo-bulge in the Milky Way.

\begin{figure}[htbp]
\sidecaption
\includegraphics[angle=0,scale=.50]{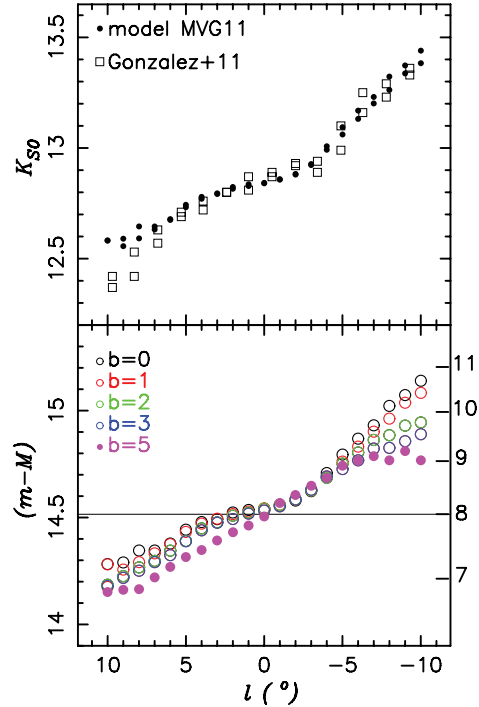}
\caption{Maxima of RC magnitude distributions in the strips
  $b=\pm1\deg$ across longitude. In the top panel for VVV K-band
  counts from \citet[][open squares]{Gonzalez+11b} and for an N-body
  model from \citet[][dots]{Martinez-Valpuesta+Gerhard11}.  The model
  reproduces the data well because its central parts near the symmetry
  plane are more axisymmetric than the main bar. In the bottom panel for
  slices through the model with different latitude as seen from the
  Sun. The change of slope seen in the simulated RC magnitude
  distributions for $|b|<2\deg$ is absent at $|b|=5\deg$. From
  \cite{Gerhard+Martinez-Valpuesta12}.}
\label{fig:3}
\end{figure}

\section{The long bar}
\label{sec:longbar}

\citet{Hammersley+00} first drew attention to an overdensity of stars
in the Milky Way disk plane reaching outwards from the bulge region to
$l\simeq 28\deg$. This structure was confirmed in several subsequent
NIR star count investigations, e.g.\ with UKIDSS
\citep{Cabrera-Lavers+08}. Due to its longitude extent and the narrow
extent along the LOS it was termed the 'long bar'. In Spitzer Glimpse
mid-infrared star counts \citep{Benjamin+05}, which are even less
affected by dust, a corresponding long bar was inferred from a similar
overdensity of sources observed on the $l>0$ side. The long bar has a
vertical scale-length of less than 100 pc, so is clearly a disk
feature. Curiously, the orientation of the long bar inferred by all
these investigations was at $\sim 45\deg$ with respect to the LOS to
the Galactic centre.

Such a misalignment of the long bar with the b/p bulge is difficult to
understand dynamically. Two independent, misaligned barred structures
with sizes differing by only a factor of $\sim 2$ would be expected to
align quickly, due to the mutual gravitational interaction. Note that
secondary nuclear bars are generally $\sim 8$ times smaller than the
primary bars in their host galaxies \citep{Erwin11}. Thus
\citet{Martinez-Valpuesta+Gerhard11, Romero-Gomez+11} both argued that
the long bar would more likely be the 2D part of the Milky Way bar
continuing outwards from the 3D barred bulge, such as generally seen
in secularly evolved N-body models.
\citet{Martinez-Valpuesta+Gerhard11} predicted star counts from an
N-body model with a bar and inner boxy bulge formed in a buckling
instability. They argued that approximate agreement with the star
count data could be achieved if the ends of the planar part of the bar
have developed leading ends through interaction with the adjacent
spiral arm heads. They found that such a configuration was present in
their model for $\sim 40\%$ of the time in an evolutionary sequence
from leading to straight to trailing, and also that some barred
galaxies have bars with leading ends.

\begin{figure}[htbp]
\sidecaption
\includegraphics[scale=.58]{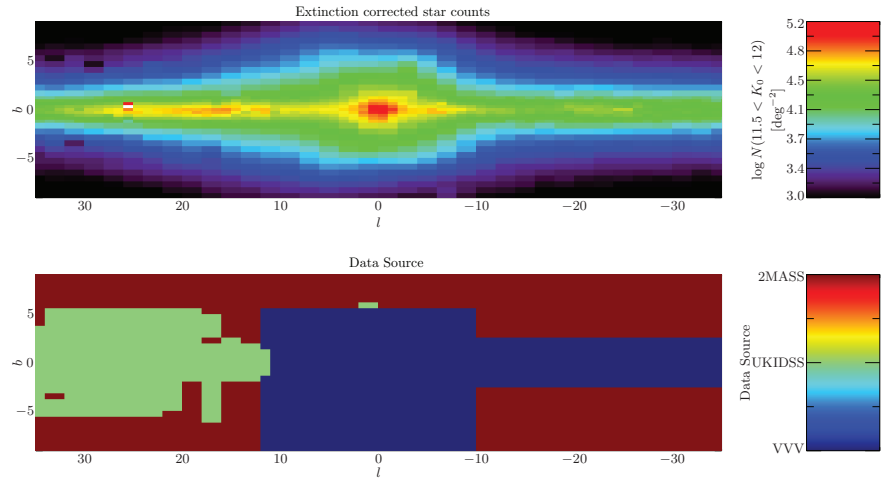}
\caption{The Galactic bulge and bar as seen from the Sun in K-band
  star counts combined from the cross-matched 2MASS, UKIDSS, and VVV
  surveys as shown in the lower panel. The magnitude range shown
  ($11.5<M_K<12$) is chosen such that the VVV data in the Galactic
  plane are not yet saturated and at the same time the 2MASS
  data are still complete in the region where they are used. This
  projection shows the continuous transition in the vertical
  scale-height from the inner 3D b/p bulge to the outer 2D bar.}
\label{fig:4}
\end{figure}

However, the star count maxima signifying the long bar
\citep{Cabrera-Lavers+08} show some discontinuities near $l=10\deg$
and $l=20\deg$, and the LOS distance dispersion of these RC stars
shows a large scatter when plotted versus longitude. Therefore, in
order to understand the Milky Way's planar bar better, we are
currently reanalyzing the combined 2MASS, UKIDSS, and VVV surveys of
the inner Galaxy (Wegg et al., in preparation). All data were
extinction corrected and brought to a common photometric
system. Fig.~\ref{fig:4} shows the total star counts across the
Galactic bulge and inner disk in a magnitude interval common to all
surveys.  From this map it is apparent that the vertical scale-height
of the star counts decreases continuously from the maximum in the
X-region of the b/p bulge into the planar bar and disk. This is
confirmed more quantitatively by measuring the vertical gradients of
the RC stars on the $l>0$ side. There is no indication for two
separate components in these plots. Investigation into the LOS
distribution of RC stars versus longitude in the combined data is
on-going.  This will give us a better understanding of the long bar
and the transition region between the bar and the adjacent disk in the
Milky Way.

\bibliographystyle{unabbr}


\end{document}